\documentclass[11pt]{article}

\usepackage[numbers,sort&compress]{natbib}
\usepackage{epsf}
\usepackage{graphicx}
\usepackage{epsfig}
\usepackage{latexsym}
\usepackage[dvips,
   bookmarksopen=true,
   bookmarksnumbered=true,
   pdftitle={Radiative energy shifts induced by local potentials},
   pdfauthor={Ulrich Jentschura},
   pdfstartview={FitB},
   colorlinks,
   linkcolor=blue]{hyperref}

\def\bbox#1{\mbox{\boldmath$#1$}}

\def\corresponds{{\lower.2ex\hbox{=}}{\rm\kern-.75em^\triangle}}
\def\succsim{\succ\kern-.9em_\sim\kern.3em}
\def\precsim{\prec\kern-1em_\sim\kern.3em}
\def\slantfrac#1#2{\kern1em^{#1}\kern-.3em/\kern-.1em_{#2}}
\def\lfrac#1#2{{}^{#1\!}\kern-.0em/_{#2}}

\def\buildrel#1\under#2{\mathrel{\mathop{\kern0pt #2}\limits_{#1}}}

\begin{document}

\bibliographystyle{myprsty}

\vspace*{0.5cm}
\begin{center}
\begin{tabular}{c}
\hline
\rule[-5mm]{0mm}{15mm}
{\Large \sf Radiative energy shifts induced by local potentials}\\
\hline
\end{tabular}
\end{center}
\vspace{0.2cm}
\begin{center}
Ulrich D. Jentschura
\end{center}
\vspace{0.2cm}
\begin{center}
{\it Theoretische Quantendynamik, \\
Fakult\"{a}t Mathematik und Physik der Universit\"{a}t Freiburg,\\
Hermann--Herder--Stra\ss{}e 3, 79104 Freiburg, Germany}
\end{center}
\vspace{0.3cm}
\begin{center}
\begin{minipage}{11.8cm}
{\underline{Abstract}}
We study a specific correction to the Bethe logarithm induced by 
potentials which are proportional to a Dirac-$\delta$ function in 
coordinate space (``local potentials''). 
Corrections of this type occur naturally in 
the calculation of various self-energy corrections to the 
energy of bound states. Examples include logarithmic 
higher-order binding corrections to the two-loop self-energy, 
vacuum-polarization induced corrections to the self-energy
and radiative corrections induced by the 
finite size of the nucleus. We obtain results for excited S and P 
states and find that the dependence of the corrections on the 
principal quantum number is remarkable.
For the ground state, we find 
a small modification as compared to previously reported
results. Our results are based on mathematical techniques for the
treatment of quantum electrodynamic bound states 
discussed previously in 
[\href{http://stacks.iop.org/JPhysA/35/1927}{J. Phys. A {\bf 35}, 1927 (2002)},
\href{http://arXiv.org/abs/hep-ph/0111084}{hep-ph/0111084}].
\end{minipage}
\end{center}
\vspace{1.3cm}

\noindent
{\underline{Journal Version:} 
\href{http://stacks.iop.org/JPhysA/36/L229}{J. Phys. A {\bf 36}, L229 (2002)}
in atomic and molecular physics,\\
quantum electrodynamics -- specific calculations,\\
relativistic and quantum electrodynamic\\
effects in atoms and molecules,\\
fine and hyperfine structure. \\

\newpage

%
%
\section{Introduction}

Ever since the quantization of the electromagnetic 
field was introduced by Dirac in~\cite{Di1927},
one of the main directions of research in the area of quantum 
electrodynamics has been the study of bound states and, notably,
the corrections to the energy of these states as induced by 
the virtual quanta. A severe limitation to the accuracy 
of current theoretical predictions is given by bound-state two-loop
self-energy effects whose evaluation has been historically 
problematic (see~\cite{JePa2002,Pa2001} and references therein).
In this Letter, we present complete results for the 
two-loop (2L) logarithmic self-energy correction of excited S states
of order
\begin{equation}
\label{defb61}
\Delta E^{(\rm 2L)}_{\rm log}(n{\rm S}) = \left( \frac{\alpha}{\pi} \right)^2 \,
\frac{(Z\alpha)^6 \, m}{n^3} \ln[(Z\alpha)^{-2}] \, B_{61}(n{\rm S})\,,
\end{equation}
where $B_{61}(n{\rm S})$ is an $n$-dependent, dimensionless
coefficient. Here, $n$ is the principal quantum number, $Z$ is the nuclear
charge number, $m$ is the electron mass,
and $\alpha$ is the fine-structure constant
(we work in natural units: $\hbar = c = \epsilon_0 = 1$).

One of the quantities which enter quite universally into 
higher-order corrections to the one- and two-loop self-energy are 
those induced by effective potentials which are proportional to
a Dirac delta-function in coordinate space. In order to calculate
$B_{61}$ for excited S states, an investigation of such corrections  
is necessary~\cite{Pa2001}. Our investigations are based
on mathematical techniques which facilitate the treatment of bound states
which rely on a separation of the virtual photons into hard (high-energy)
and soft (low-energy) quanta; these have recently been generalized to 
two-loop effects~\cite{JePa2002,JeKePa2002}.

%
%
\section{General Formulation}

Let us consider the Schr\"{o}dinger Hamiltonian
\begin{equation}
H = \frac{\bbox{p}^2}{2 m} + V\,,
\end{equation}
where $V$ is the binding Coulomb potential (the energy 
eigenvalues are $E = -(Z\alpha)^2 \,m/(2 n^2)$).
We assume a small perturbation of $H$ 
proportional to a $\delta$-function,
\begin{equation}
\label{deltaV}
\delta V = \frac{\pi \, (Z\alpha) \, \delta^{(3)}(\bbox{r})}{m^2}\,.
\end{equation}
For S states, this perturbation leads to an energy shift
\begin{equation}
\label{deltaE}
\delta E = \langle \phi | \delta V | \phi \rangle \,,
\end{equation}
where $|\phi\rangle$ is the electron wave function for which
the nonrelativistic approximation may be used in the context of the 
evaluation of the radiative corrections as discussed in this Letter.
The correction (\ref{deltaE})
is nonvanishing only for S states, where the correction
to the $n{\rm S}$-state energy amounts to $(Z\alpha)^4/n^3$.
The perturbation (\ref{deltaV}) also induces a modification of the
wave function
\begin{equation}
\label{deltaphi}
| \delta \phi \rangle = \left( \frac{1}{E-H} \right)' \, 
\delta V | \phi \rangle\,,
\end{equation}
where the prime denotes the reduced Green function (in the spectral 
decomposition, the reference state $|\phi\rangle$ is excluded).

We calculate the delta-like correction
to the one-loop Bethe logarithm,
\begin{eqnarray}
\label{Cdelta}
\Delta E^{(\rm L)}_\delta(nl,\overline{\epsilon}) &=& \frac{2\alpha}{3\pi} \,
\delta_{\delta V} \, \left\{
\int_0^\epsilon {\rm d}\omega\,\omega\,
\left< \phi \left| \frac{p^i}{m} \,
\frac{1}{E - (H + \omega)} \, 
\frac{p^i}{m} \right| 
\phi \right> \right\} \nonumber\\[2ex]
&=& \frac{\alpha}{\pi} \, (Z\alpha)^6 \,
\frac{m}{n^3} \, F_6(nl,\overline{\epsilon})\,,
\end{eqnarray}
where $nl$ denotes the bound-state quantum numbers.
The notation (\ref{Cdelta})
follows the usual spectroscopic nomenclature, $n$ is the 
principal quantum number, and $l$ is the orbital angular 
momentum. The correction (\ref{Cdelta})
is independent of the electron spin, and $\overline{\epsilon}$
denotes the quantity
\begin{equation}
\label{defeps}
\overline{\epsilon} = \frac{\epsilon}{(Z\alpha)^2}
\end{equation}
(the notation has been introduced in Ref.~\cite{Pa2001}).
The upper index (L) in Eq.~(\ref{Cdelta}) is assigned because
$\Delta E^{(\rm L)}_\delta$ represents the low-energy part of the 
correction (due to soft virtual photons), in the language of the
formalism introduced in~\cite{Pa1993,JePa1996,JePa2002}.
In Eq.~(\ref{Cdelta}), the symbol $\delta_{\delta V}$,
inspired by the notation of~\cite{Pa2001}, denotes the first-order
perturbation received by the quantity in curly brackets through the
replacements [see Eqs.~(\ref{deltaV}), (\ref{deltaE}), (\ref{deltaphi})] 
\begin{eqnarray}
H &\to& H + \delta V\,, \\[2ex]
|\phi\rangle &\to& |\phi\rangle + |\delta \phi\rangle\,, \\[2ex]
E &\to& E + \delta E\,. 
\end{eqnarray}
The quantity $F_6(nl,\overline{\epsilon})$ in Eq.~(\ref{Cdelta})
is a dimensionless function which parametrizes an
effect of the order of $(Z\alpha)^6$.
Here, we present results for $n{\rm S}$ states in the
range $n=1,\dots,8$ and $n{\rm P}$ states ($n=2,\dots,8$).

%
%
\begin{figure}[htb]
\begin{center}
\begin{minipage}{12cm}
\centerline{\mbox{\epsfysize=4.5cm\epsffile{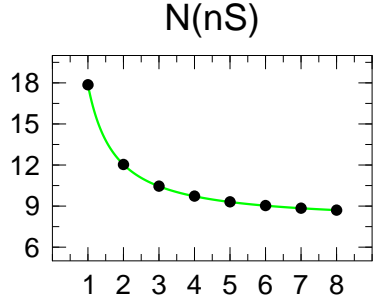}}\hbox to 0.75in{}}
\caption{\label{fig1} The dependence of the
correction $N(n{\rm S})$ on the principal quantum number
is remarkable. For $n=8$, the magnitude of the 
result is less than half $N(1{\rm S})$.
The smooth curve is a fit based on a model based on a three-parameter
fit of the form $a_{\rm S} + b_{\rm S}/n + c_{\rm S}/n^2$ 
[see Eq.~(\ref{fitnS})].}
\end{minipage}
\end{center}
\end{figure}

%
%
\section{Correction to S states}

For $S$ states, $F_6$ as implicitly defined in Eq.~(\ref{Cdelta})
has the form~\cite{Pa2001}
\begin{eqnarray}
\label{F6nS}
\label{defNS}
F_6(n{\rm S},\overline{\epsilon}) &=& -\frac23 \, \ln^2 (\overline{\epsilon})
+ \ln (\overline{\epsilon}) \, \biggl[ 2 \, \{ 1 - \ln 2 \}
\nonumber\\[2ex]
& & + \frac83 \, \left( \frac34 + \frac{1}{4 n^2} - \frac1n - \ln(n) +
\Psi(n) + C \right) \biggr] + N(n{\rm S}) \,.
\end{eqnarray}
Here, the notation $\overline{\epsilon}$ is explained in
Eq.~(\ref{defeps}),
and $N(n{\rm S})$ is a nonlogarithmic term which has been 
known only for $n=1$ (see Ref.~\cite{Pa2001}). The dependence on 
$\overline{\epsilon}$ cancels when the high-energy part is added
to the above result~\cite{Pa1993,JePa1996,Pa1996}. The constant 
term $N(n{\rm S})$
has been shown to contribute to the two-loop self-energy coefficient
$B_{61}(n{\rm S})$ [see \cite[Eq.~(50)]{Pa2001}]. 
The complexity of the 
calculation increases with increasing principal quantum
number, because of the more complex structure of the bound-state
wave function and the necessity to subtract poles of the
integrand corresponding to the decay into 
lower-energy states. For states with $n=8$, we obtain intermediate
results with $198,000$ terms, and use is made of computer 
algebra systems~\cite{Wo1988}.

We obtain the following
results for $N(n{\rm S})$ is the range $n=1,\dots,8$:
\begin{eqnarray}
\label{resultsnS}
N(1S) &=& 17.855\,672(1)\,, \quad
N(2S) = 12.032\,209(1)\,, \nonumber\\[2ex]
N(3S) &=& 10.449\,810(1)\,, \quad 
N(4S) = 9.722\,413(1)\,, \nonumber\\[2ex]
N(5S) &=& 9.304\,114(1)\,, \quad
N(6S) = 9.031\,832(1)\,, \nonumber\\[2ex]
N(7S) &=& 8.840\,123(1)\,, \quad
N(8S) = 8.697\,639(1)\,. 
\end{eqnarray}
A least-squares fit with a functional form
\begin{equation}
\label{modelnS}
N(n{\rm S}) \approx a_{\rm S} + b_{\rm S}/n + c_{\rm S}/n^2
\end{equation}
yields the fit-parameter values
\begin{equation}
\label{fitnS}
a_{\rm S} = 7.78\,,\quad
b_{\rm S} = 3.13\,,\quad
c_{\rm S} = 6.93\,.
\end{equation}
The data in Eq.~(\ref{resultsnS}) are well
represented by this fit, as is evident from Fig.~\ref{fig1}.
The excellent agreement between the fit and the 
numerical data in Eq.~(\ref{resultsnS}) could suggest
a finite limit
\begin{equation}
\lim_{n\to\infty} N(n{\rm S}) \approx 7.78\,.
\end{equation}
The functional form of the fit is inspired, in particular, 
by the term proportional to $\ln(\overline{\epsilon})$
in (\ref{F6nS}). This term can be expanded in a series involving
inverse powers of $n$. Note that the logarithmic term $\ln (n)$,
for large $n$, cancels against a compensating term originating from the 
expansion of $\Psi(n)$.

%
%
\section{Correction to P states}

For P states, $F_6(n{\rm P},\overline{\epsilon})$ assumes the functional form
\begin{eqnarray}
\label{F6nP}
F_6(n{\rm P},\overline{\epsilon}) &=&  
\frac{2}{9} \, 
\left( 1 - \frac{1}{n^2} \right) \, \ln \overline{\epsilon}
+ N(n{\rm P}) \,.
\end{eqnarray}
We obtain the results,
\begin{eqnarray}
\label{resultsnP}
N(2{\rm P}) &=& 0.003\,300\,635(1)\,, \quad
N(3{\rm P}) = 0.003\,572\,084(1)\,, \nonumber\\[2ex]
N(4{\rm P}) &=& -0.000\,394\,332(1)\,, \quad
N(5{\rm P}) = -0.004\,303\,806(1)\,, \nonumber\\[2ex]
N(6{\rm P}) &=& -0.007\,496\,998(1)\,, \quad
N(7{\rm P}) = -0.010\,014\,614(1)\,, \nonumber\\[2ex]
N(8{\rm P}) &=& -0.011\,999\,223(1)\,.
\end{eqnarray}
The result for $N(2{\rm P})$ was previously obtained 
in~\cite[Eq.~(4.159)]{Je1996}. Specifically, the contribution due to this term
entered into the result for $F_{\delta H}$ given in~\cite[Table I]{JePa1996}.
Here, we present the generalization of the result to higher principal
quantum numbers. Observe that the correction changes its sign as $n$ is 
increased. In analogy to the $S$ states, we use a least-squares fit 
with a functional form
\begin{equation}
N(n{\rm P}) \approx a_{\rm P} + b_{\rm P}/n + c_{\rm P}/n^2\,,
\end{equation}
which results in the fit-parameter values
\begin{equation}
\label{fitnP}
a_{\rm P} = -0.030\,,\quad
b_{\rm P} = 0.170\,,\quad
c_{\rm P} = -0.206\,.
\end{equation}
The excellent agreement between the fit and the numerical
data is represented in Fig.~\ref{fig2}.

%
%
\begin{figure}[htb]
\begin{center}
\begin{minipage}{12cm}
\centerline{\mbox{\epsfysize=4.5cm\epsffile{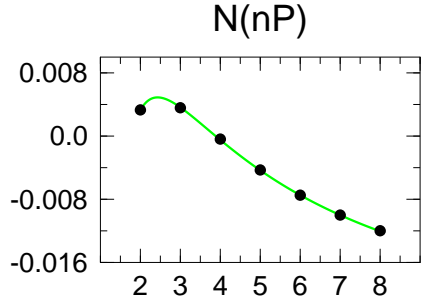}}\hbox to 0.75in{}}
\caption{\label{fig2} The analogue of
Fig.~\ref{fig1} for P states.}
\end{minipage}
\end{center}
\end{figure}

The spin-dependent high-energy part for P states is
\begin{equation}
\Delta E^{(\rm H)}_\delta(n{\rm P}_j,\epsilon) = 
\frac{\alpha}{\pi} \, (Z\alpha)^6 \,
\frac{m}{n^3} \, H_6(n{\rm P}_j,\overline{\epsilon})\,.
\end{equation}
Inspired by the effective treatment of radiative corrections 
based on form factors~\cite[Ch.~7]{ItZu1980}, we take into account
the contribution due to hard virtual photons by the replacement
\begin{equation}
\label{formfactor1}
\delta V \to \delta V \, F_1(-\bbox{q}^2)\,,
\end{equation}
but this does not give the complete result. 
We also have to consider the perturbation of the spin-dependent 
interaction
\begin{equation}
\label{formfactor2}
\frac{1}{2 m}\, \left( {\rm i}\,\bbox{\gamma} \cdot \delta \bbox{E} \right) \to
\frac{1}{2 m}\, \left( {\rm i}\,\bbox{\gamma} \cdot \delta \bbox{E} \right)
\, F_2(-\bbox{q}^2)\,,
\end{equation}
where $\delta \bbox{E}$ is the electric field generated by the 
perturbating potential $\delta V$, i.e.
\begin{equation}
\delta \bbox{E} = -{\rm i}\, \bbox{q} \, \frac{\pi (Z\alpha)}{m^2}
\end{equation}
in momentum space, and $\bbox{\gamma}$ is a three-vector 
whose elements are the spatial Dirac matrices~\cite{ItZu1980}. 
Both of the above replacement prescriptions 
(\ref{formfactor1}) and (\ref{formfactor2}) are dictated 
by the modified Dirac Hamiltonian as given in \cite[Eq.~(3)]{JePa2002}. 

We first consider the correction due to the charge form factor
$F_1$ as given in Eq.~(\ref{formfactor1}). Because we consider {\em a priori}
only the one-loop effect, we may take $F_1$ in the one-loop approximation
and expand only up to the order of $\bbox{q}^2$. A suitably chosen
perturbating potential may be used for an effective treatment
of further loops, as discussed in Sec.~\ref{Conclusions} below.
Formulas relevant to the electron charge form factor $F_1$ 
can be found in Eqs.~(5), (10), and (31) of~\cite{JePa2002}. We find
\begin{equation}
\Delta E^{(\rm H,1)}_\delta(n{\rm P}_j,\epsilon) = 
\alpha \, \frac{Z\alpha}{3 m^4} \,
\left( \ln \left(\frac{m}{2\,\epsilon}\right) + \frac{11}{24} \right) \,
\left.\Delta(|\psi_{n{\rm P}}(\bbox{r})|^2)\right|_{\bbox{r} = \bbox{0}}\,,
\end{equation}
where $\psi_{n{\rm P}}(\bbox{r})$ is the nonrelativistic wave function
of the $n{\rm P}$ state, and the matrix element reads
\begin{equation}
\left.\Delta(|\psi_{n{\rm P}}(\bbox{r})|^2)\right|_{\bbox{r} = \bbox{0}} =
\frac23 \, \frac{(Z\alpha)^5\,m^5}{\pi \, n^3} \,
\left( 1 - \frac{1}{n^2} \right) \,.
\end{equation}
Therefore, the result for the scaled high-energy part due to 
the $F_1$ form factor reads
\begin{equation}
\label{H6nP1}
H^{(1)}_6(n{\rm P}_j,\epsilon) = 
\frac{2}{9} \, \left( 1 - \frac{1}{n^2} \right) \,
\left( \ln \left(\frac{m}{2\,\epsilon}\right) + \frac{11}{24} \right)\,.
\end{equation}

Now we turn to the spin-dependent correction proportional 
to the magnetic form factor $F_2$ as given in Eq.~(\ref{formfactor2}).
Again, we may employ the one-loop approximation and take
$F_2$ at zero momentum, where its well-known Schwinger value reads
$F_2(0) = \alpha/(2\pi)$. We find
\begin{equation}
\Delta E^{(\rm H,2)}_\delta(n{\rm P}_j,\epsilon) =
F_2(0) \, \frac{\pi \,(Z\alpha)}{2 m^3} \, 
\left< n{\rm P}_j \left| \bbox{\gamma}\cdot \bbox{q} \right| 
n{\rm P}_j \right> \,,
\end{equation}
where the expectation value of the momentum space operator 
$\bbox{\gamma}\cdot \bbox{q}$ reads
\begin{equation}
\left< n{\rm P}_j \left| \bbox{\gamma}\cdot \bbox{q} \right| n{\rm P}_j \right> 
= {\mathrm i} \, \left[ \frac{\partial}{\partial \bbox{x}}
\left(\psi_{n{\rm P}_j}^{+}(\bbox{r}) \bbox{\gamma} 
\psi_{n{\rm P}_j}(\bbox{r}) \right)
  \right]_{r=0}
=  \frac{n^2 - 1}{\pi\,n^5} \, (Z\alpha)^5\,m^4\, \delta_{j,1/2}\,,
\end{equation}
i.e. the matrix element vanishes for ${\rm P}_{3/2}$ states
[please observe the missing factor $\pi$ in the 
denominator of the right-hand side of Eq.~(29) of~\cite{JePa2002}]. 
We thus obtain
\begin{equation}
\label{H6nP2}
H^{(2)}_6(n{\rm P}_j,\epsilon) =
\frac{1}{4} \, \left( 1 - \frac{1}{n^2} \right) \,
\delta_{j,1/2}
\end{equation}
in analogy to the result obtained in Eq.~(30) of 
Ref.~\cite{JePa2002}.

Adding the low-energy part (\ref{F6nP}), and the two
high-energy contributions (\ref{H6nP1}) and (\ref{H6nP2}) 
due to $F_1$ and $F_2$, respectively, we obtain
\begin{equation}
\label{EdeltanP}
\Delta E_\delta(n{\rm P}_j) =
\Delta E^{(\rm L)}_\delta(n{\rm P},\overline{\epsilon}) +
\sum_{k=1,2} \Delta E^{(\rm H,k)}_\delta(n{\rm P}_j,\epsilon) =
\frac{\alpha}{\pi} \, (Z\alpha)^6 \,
\frac{m}{n^3} \, F_\delta(n{\rm P}_j)\,,
\end{equation}
where the function $F_\delta(n{\rm P})$ is independent of $\epsilon$
and has the form 
\begin{eqnarray}
\label{FdeltanP}
F_\delta(n{\rm P}_j) &=& F_6(n{\rm P},\overline{\epsilon}) + 
\sum_{k=1,2} H^{(k)}_6(n{\rm P}_j,\epsilon) \nonumber\\[2ex]
&=& \frac{2}{9} \, \left( 1 - \frac{1}{n^2} \right) \,
\left( \ln \left(\frac{1}{(Z\alpha)^2}\right) - \ln (2) + 
\frac{11}{24} + \frac{9}{8} \, \delta_{j,1/2} \right) + N(n{\rm P})\,.
\end{eqnarray}
The numerical values of $N(n{\rm P})$ are given in Eq.~(\ref{resultsnP}).

%
%
\begin{figure}[htb]
\begin{center}
\begin{minipage}{12cm}
\centerline{\mbox{\epsfysize=5.5cm\epsffile{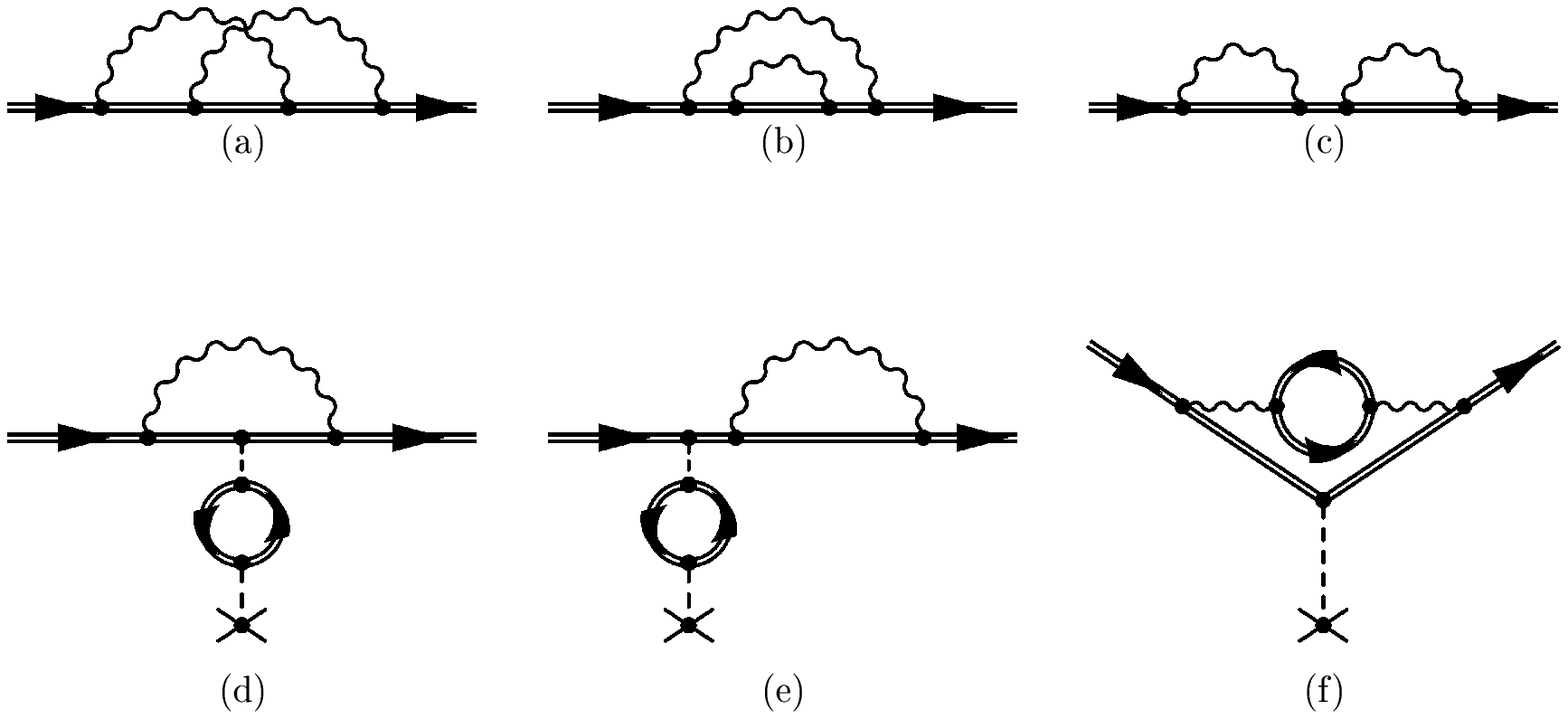}}\hbox to 0.75in{}}
\caption{\label{fig3} Some of 
two-loop diagrams involving the bound-state electron 
self-energy which contribute to the logarithmic coefficient
$B_{61}$. All of these (and others) were taken into account
in Ref.~\cite{Pa2001}. The double line in the diagrams
denotes the bound electron (Dirac--Coulomb propagator).
Figs.~(a)--(c) contribute also to the 
triple logarithm of order $\alpha^2\,(Z\alpha)^6\,\ln^3[(Z\alpha)^{-2}]\,m$,
i.e.~to the $B_{63}(n{\rm S})$-coefficient.
The double logarithm of order $\alpha^2\,(Z\alpha)^6\,\ln^2[(Z\alpha)^{-2}]\,m$
originates from diagrams (a)--(e). General formulas valid for 
$B_{62}(n{\rm S})$ and $B_{63}(n{\rm S})$ 
of arbitrary $n$ are given in Eqs.~(48), (49) and~(51)
of Ref.~\cite{Pa2001}. The single logarithm $B_{61}(n{\rm S})$
involves many more contributions, for example those originating from the 
diagram (f). This Feynman diagram is considered in Eq.~(40) of~\cite{Pa2001},
together with contributions from other diagrams involving vacuum
polarization effects. Complete results for $B_{61}(n{\rm S})$ are given
in Eq.~(\ref{b61results}).}
\end{minipage}
\end{center}
\end{figure}

%
%
\section{Conclusions}
\label{Conclusions}

We summarize the results of this Letter.
\begin{itemize}
\item 
{\bf Dependence of the $B_{61}$ coefficient on the principal quantum number.}
Currently, one of the most important limiting factors to a 
further progress of the theory of the S state Lamb shift is the 
understanding of higher-order binding corrections to the two-loop 
bound-state self-energy~\cite{MoTa2000}. 
The theory of S states is important for the 
deduction of the Rydberg constant, and it is also necessary to 
investigate higher excited states because the frequencies of more
than one transition have to be theoretically known in order to 
infer the fundamental constants (see~\cite{MoTa2000} and references therein). 

Our result for $N(1{\rm S})$ in Eq.~(\ref{resultsnS}) differs slightly from 
previously published results [see Eq.~(21) of~\cite{Pa2001}].
General formulas, whose structure is valid
for $B_{61}(n{\rm S})$ of arbitrary $n$
have been given in Ref.~\cite[Eqs.~(50) and~(52)]{Pa2001};
these are relevant to the sum of
two-loop self-energy {\em and} vacuum-polarization effects. 
Some of the contributing Feynman diagrams are shown in Fig.~\ref{fig3}.
However, the quantity $N(n{\rm S})$ which enters into the expression
for $B_{61}(n{\rm S})$ has been known only for $n=1$.
Here, we take into account the 
slightly shifted valued of $N(1{\rm S})$ as well as the results
presented in Eq.~(\ref{resultsnS}) for $n > 1$.
We finally obtain the following results for $B_{61}(n{\rm S})$
defined in Eq.~(\ref{defb61}),
\begin{eqnarray}
\label{b61results}
B_{61}(1{\rm S}) &=& 50.344\,005(1) \,,\quad
B_{61}(2{\rm S}) = 42.447\,669(1) \,,\nonumber\\[2ex]
B_{61}(3{\rm S}) &=& 40.289\,637(1) \,,\quad
B_{61}(4{\rm S}) = 39.294\,929(1) \,,\nonumber \\[2ex]
B_{61}(5{\rm S}) &=& 38.722\,048(1) \,,\quad
B_{61}(6{\rm S}) = 38.348\,805(1) \,,\nonumber\\[2ex]
B_{61}(7{\rm S}) &=& 38.085\,860(1) \,,\quad
B_{61}(8{\rm S}) = 37.890\,354(1) \,.
\end{eqnarray}
This completes the calculation of logarithmic two-loop
corrections to S states of order $\alpha^2\,(Z\alpha)^6\,
\ln^j[(Z\alpha)^{-2}]\,m$ ($j=1,2,3$).
\item
{\bf Clarification of a specific intermediate step
in the calculation of the muonium hyperfine splitting.}
In lowest order, the hyperfine splitting Hamiltonian for an electron
bound to an atomic nucleus with magnetic moment $\bbox{\mu}_n$ is 
given by~\cite[p.~79]{ItZu1980}:
\begin{eqnarray}
\label{Vhfs}
V_{\rm hfs}(\bbox{r}) &=& \frac{e}{2\,m}\, \bbox{\sigma}_{\rm e}\,\cdot\,
\left[ \bbox{\nabla} \times (\bbox{\mu}_n \times \bbox{\nabla}\right] \,
\frac{1}{4\,\pi\,r} \nonumber\\[2ex]
&=& \frac{e}{8\,\pi\,m}\, \bbox{\sigma}_{\rm e}\,\cdot\,
\left[ \bbox{\mu}_n \, \bbox{\nabla}^2 -
(\bbox{\mu}_n \cdot \bbox{\nabla}) \, \bbox{\nabla} \right] \,
\frac{1}{r} \nonumber\\[2ex] 
&\to&  -\frac{e}{3\,m} \, 
\left(\bbox{\sigma}_{\rm e}\,\cdot\, \bbox{\mu}_n\right) \,
\delta^{(3)}(\bbox{r})\,,
\end{eqnarray}
which is a delta-like potential as in (\ref{deltaV}). The angular averaging 
$\nabla_i\,\nabla_j \to (1/3)\, \delta_{ij} \,
\bbox{\nabla}^2$ [see the transition from the second to the third
line of Eq.~(\ref{Vhfs})] is valid when $V_{\rm hfs}(\bbox{r})$
is evaluated on S states. Therefore, the quantity $N(n{\rm S})$
appears naturally in the evaluation of radiative corrections to the 
hyperfine splitting of S states, for example in muonium~\cite{Pa1996,NiKi1997}.

The results for $N(1S)$ given in Eq.~(\ref{resultsnS}) differs slightly
from the corresponding previously published value for the low-energy
part of the muonium hyperfine splitting 
as given in~\cite{Pa1996}. The difference is
\begin{equation}
\frac{2}{3} \, \left(\frac{31}{36} - \frac{\pi^2}{12}\right) = 
0.0257627\dots 
\end{equation}
While the new result reported here
does not affect the final result for the hyperfine splitting,
it explains the discrepancies between the
intermediate results of~\cite{Pa1996} and those of Ref.~\cite{NiKi1997}.
\item
{\bf Vacuum-polarization induced correction to the 
self-energy.} Our results for $N(n{\rm P})$ in Eq.~(\ref{resultsnP})
and the total results in Eqs.~(\ref{EdeltanP}) and (\ref{FdeltanP})
for the energy shift due to the radiative correction
to a delta-like potential can be
used in order to evaluate the contribution to the 
energy of a P state due to the diagram in Fig.~\ref{fig3} (d).
This is a combined ``self-energy vacuum-polarization'' 
correction for P states generated by a
vacuum polarization correction to the Coulomb exchange in the 
bound electron propagator within the one-loop self-energy
diagram. The lowest-order one-loop
vacuum-polarization potential reads~\cite[p.~327]{ItZu1980}
\begin{equation}
V_{\rm V.P.}(\bbox{r}) = -\frac{4}{15} \,  
\frac{\alpha}{\pi} \,
\frac{\pi \, (Z\alpha) \, \delta^{(3)}(\bbox{r})}{m^2}\,,
\end{equation}
which is proportional to $\delta V$ defined in Eq.~(\ref{deltaV}).
Our results in Eqs.~(\ref{resultsnP}),
(\ref{EdeltanP}) and (\ref{FdeltanP}) contribute to the 
two-loop coefficients $B_{61}$ and $B_{60}$ for P states as
defined in~\cite{JePa2002}.
\item
{\bf Nuclear finite-size correction to the self-energy.} 
For an atomic nucleus whose root-mean-square charge radius is small
compared to the Bohr radius, the effect of the nuclear finite size 
can be incorporated as a form-factor correction to the Coulomb
interaction in full analogy to Eqs.~(\ref{formfactor1})
and~(\ref{formfactor2}).
In this case the appropriate form factor reads
\begin{equation}
F(-\bbox{q}^2) = 1 - \bbox{q}^2 \,
\frac{\langle \bbox{r}^2 \rangle}{6}
\end{equation}
where $\langle \bbox{r}^2 \rangle$ is the mean square 
radius of the charge distribution of the atomic nucleus.
The Dirac delta-like finite-size ``potential'' therefore reads 
\begin{equation}
V_{\rm f.s.} = \frac{2}{3}\, \langle \bbox{r}^2 \rangle \,
\frac{\pi \, (Z\alpha) \, \delta^{(3)}(\bbox{r})}{m^2}\,,
\end{equation}
which is again proportional to $\delta V$ as defined in Eq.~(\ref{deltaV}).
Therefore, the results in this Letter can be used for an evaluation 
of the finite-size correction to the self-energy in atomic systems with a 
low nuclear charge number. Our result in Eq.~(\ref{resultsnP})
for the case $n=2$ has been obtained previously
in~\cite{Je1996}. For the $2{\rm P}_{1/2}$ and $2{\rm P}_{3/2}$ states,
our calculations are also in agreement with Eqs.~(16) and 
(17) of Ref.~\cite{MiSuTe2002}.
Our results in Eqs.~(\ref{resultsnP}) and (\ref{FdeltanP}) above 
represent generalizations of this previous
work to higher excited states.
\end{itemize}
Work on the nonlogarithmic term $B_{60}$ for S states
is currently in progress and will be presented elsewhere, 
together with improved theoretical values for the Lamb shift of S states
as derived from the results presented in this Letter.

\section*{Acknowledgements}

Insightful discussions with K. Pachucki are gratefully acknowledged.

\end{document}